\documentclass[%
 aip,
 amsmath,amssymb,
reprint,
]{revtex4-1}

\usepackage{graphicx}
\usepackage{subfigure}
\usepackage{dcolumn}
\usepackage{bm}

\usepackage[utf8]{inputenc}
\usepackage[T1]{fontenc}
\usepackage{mathptmx}
\usepackage{etoolbox}

\usepackage{url}
\usepackage[table,xcdraw]{xcolor}

\usepackage[pdfauthor={Karen Fidanyan, Mariana Rossi},
            pdftitle={Manuscript}]{hyperref} 
\hypersetup{colorlinks,
    citecolor=black,
    filecolor=black,
    linkcolor=black,
    urlcolor=black
}
\usepackage{tabularx}
\newcolumntype{L}{>{$}l<{$}} 
\newcolumntype{R}{>{\raggedleft\arraybackslash}X}

\newcommand{\rev}[1]{\textcolor{black}{#1}}

\usepackage[acronym,shortcuts]{glossaries}
\makenoidxglossaries

\newacronym{aimd}{AIMD}{\textit{ab initio} molecular dynamics}
\newacronym{aipimd}{AI-PIMD} {\textit{ab initio} path integral molecular dynamics}
\newacronym{bfgs}{BFGS}{Broyden–Fletcher–Goldfarb–Shanno}
\newacronym{bo}{BO}{Born-Oppenheimer}
\newacronym{cineb}{CI-NEB} {climbing-image nudged elastic band}
\newacronym{dft}{DFT} {density-functional theory}
\newacronym{dof}{DOF}{degree of freedom}
\newacronym{dos}{DOS}{density of states}
\newacronym{esm}{ESM}{effective screening medium}
\newacronym{fbz}{FBZ}{first Brillouin zone}
\newacronym{gcscf}{GCSCF}{grand-canonical self-consistent field}
\newacronym{gf}{GF} {Green's function}
\newacronym{gga}{GGA}{generalized gradient approximation}
\newacronym{gle}{GLE}{Generalized Langevin eqiation}
\newacronym{ha}{HA}{harmonic approximation}
\newacronym{heg}{HEG}{homogeneous electron gas}
\newacronym{homo}{HOMO}{highest occupied molecular orbital}
\newacronym{hse06}{HSE06}{Heyd–Scuseria–Ernzerhof-2006}
\newacronym{ksdft}{KS-DFT}{Kohn-Sham density-functional theory}
\newacronym{lda}{LDA}{local density approximation}
\newacronym{leed}{LEED}{low energy electron diffraction}
\newacronym{lzk}{LZK}{Lifshitz-Zaremba-Kohn model}
\newacronym{mbd}{MBD}{many-body dispersion}
\newacronym{mbd-rsscs}{MBD@rsSCS}{many-body dispersion with range-separated self-consistent screening}
\newacronym{mc}{MC}{Monte-Carlo}
\newacronym{md}{MD}{molecular dynamics}
\newacronym{mep}{MEP}{minimal-energy path}
\newacronym{neb}{NEB}{nudged elastic band}
\newacronym{nlmbd}{nl-MBD}{non-local many-body dispersion}
\newacronym{nqe}{NQE}{nuclear quantum effects}
\newacronym{pbc}{PBC}{periodic boundary conditions}
\newacronym{pbe}{PBE}{Perdew–Burke-Ernzerhof}
\newacronym{pes}{PES}{potential energy surface}
\newacronym{piglet}{PIGLET}{path integrals with generalized Langevin equation}
\newacronym{pimd}{PIMD}{path integral molecular dynamics}
\newacronym{pl}{PL}{principal layer}
\newacronym{slrpc}{SL-RPC}{spatially localized ring polymer contraction}
\newacronym{stm}{STM}{scanning tunneling microscopy}
\newacronym{qh}{QH}{quasi-harmonic}
\newacronym{qha}{QHA}{quasi-harmonic approximation}
\newacronym{qho}{QHO}{quantum harmonic oscillator}
\newacronym{scf}{SCF}{self-consistent field}
\newacronym{sfg}{SFG}{sum-frequency generation spectroscopy}
\newacronym{sie}{SIE}{self-interaction error}
\newacronym{tpd}{TPD}{temperature-programmed desorption}
\newacronym{tst}{TST}{transition state theory}
\newacronym{ups}{UPS}{ultraviolet photoemission spectroscopy}
\newacronym{vdos}{vDOS}{vibrational density of states}
\newacronym{vdw}{vdW}{van der Waals}
\newacronym{xc}{XC}{exchange-correlation}
\newacronym{zpe}{ZPE}{zero point energy}

\makeatletter
\def\@email#1#2{%
 \endgroup
 \patchcmd{\titleblock@produce}
  {\frontmatter@RRAPformat}
  {\frontmatter@RRAPformat{\produce@RRAP{*#1\href{mailto:#2}{#2}}}\frontmatter@RRAPformat}
  {}{}
}%
\makeatother
\begin{document}

\preprint{PREPRINT}

\title{Ab initio study of water dissociation on a charged Pd(111) surface}

\author{Karen Fidanyan}
\affiliation{Max Planck Institute for the Structure and Dynamics of Matter, Luruper Chaussee 149, 22761 Hamburg, Germany}

\author{Guoyuan Liu}
\affiliation{Department of Materials Science and Engineering, École Polytechnique Fédérale de Lausanne, CH-1015 Lausanne, Switzerland}

\author{Mariana Rossi}
\email{mariana.rossi@mpsd.mpg.de}
\affiliation{Max Planck Institute for the Structure and Dynamics of Matter, Luruper Chaussee 149, 22761 Hamburg, Germany}
\affiliation{Fritz Haber Institute of the Max Planck Society, Faradayweg 4-6, 14195 Berlin, Germany}

\begin{abstract}
Interactions between molecules and electrode surfaces play a key role in electrochemical processes and are a subject of extensive research, both experimental and theoretical. 
In this manuscript, we address the water dissociation reaction on a Pd(111) electrode surface, modelled as a slab embedded in an external electric field. We aim at unraveling the relationship between surface charge and zero-point-energy in aiding or hindering this reaction. We calculate energy barriers with dispersion-corrected density-functional theory and an efficient parallel implementation of the nudged-elastic-band method.
We show that the lowest dissociation barrier, and consequently highest reaction rate, takes place when the field reaches a strength \rev{where two different geometries of the water molecule in the reactant state are equally stable}. Zero-point energy contributions to this reaction, on the other hand, remain nearly constant across a wide range of electric field strengths, despite significant changes in the reactant state. \rev{Interestingly, we show that the application of  electric fields that induce a negative charge on the surface can make nuclear tunneling more significant for these reactions.}
\end{abstract}

\maketitle
\newpage 

\section{Introduction} \label{sec_intro}

Surface-mediated reactions involving water appear in many technologically relevant processes that lead to hydrogen production or material corrosion~\cite{wuerger_scirep_2020}, for example. In particular in electrocatalysis -- a key process to achieve a sustainable economy -- the chemistry involved in the water splitting reaction can be \rev{modified by} the application of a potential bias~\cite{NongJones2020}.
In such a system, the structure of water at the interface, the collective properties of the surface, and the electrolyte composition affect the reactions substantially, and present a challenge both for theory and experiments, even when considering idealized electrodes~\cite{WangFeng2021, Tong_2017_expt_water_at_electrode, Toney_1994_voltage_dep_ordering_expt}.

From a theoretical perspective, there are at least three ingredients that must be accounted for when addressing water splitting at (charged) metallic surfaces: charge transfer and reorganization at the interface, (screened) van der Waals interactions between molecules and metal, and the effect of nuclear fluctuations and dynamics on the process of interest. It should be noted that the latter is now accepted to be adequately described only by considering the nuclei as quantum particles~\cite{Laasonen_Car_1993_ab_initio_water, Marx_1999_AIPIMD_water, Marx_2000_water,  LiMichaelides2010, Rossi:2014gq, Ceriotti_Michaelides_2016_NQE_in_water}.
While the structure of water at metallic surfaces has been addressed by theoretical models numerous times in the past~\cite{Carrasco_Michaelides_2013_role_of_vdW, Izvekov_2001_water_Ag111_AIMD, Gross_Sakong_2022_Ab_initio_water_interfaces}, even including potential biases in the simulations~\cite{Gross_Sakong_2022_Ab_initio_water_interfaces,Pedroza_2015_liquid_water, GoldsmithSelloniChemSci2021,Sugino_2007_bias_water}, the surface-mediated dissociation of water has received comparatively less attention from theoretical work \rev{despite its undeniable importance}.
\rev{For example, in alkaline conditions, the hydrogen dissociation of a water molecule (Volmer step) is the first step of the hydrogen evolution reaction and it is likely to be rate-determining in these conditions\cite{DubouisGrimaud2019}.}
Several groups reported studies of the splitting of water molecules on various transition metals~\cite{Wang_Tao_Bu_2006_water_dissoc_ANEBA, Cao_2006_water_dissoc_CINEB, Phatak_2009_water_dissoc_CINEB, Fajin_2010_water_dissoc_DIMER, Donadio:2012jq, Litman_2018_SLRPC, Qian_Hamada_2013_water_Pt111_bias}, but to the best of our knowledge, never accounting at the same time for potential biases, long-range dispersion interactions and \gls*{nqe}. 

Specifically, accounting for a potential bias in first-principles simulation is not trivial.
A number of methods to include a potential bias in the slab model were proposed, including various degrees of approximation,~\cite{Otani_Sugino_2006, Bonnet_Otani_2012_Potentiostat, Sundararaman_2017_GCSCF, Surendralal_2018, Hagiwara_Otani_2021, Deissenbeck_2021_canon_potstat, KhatibCucinotta2021, AhrensMeissnerElectrode2022} without the adoption of a common ``default'' approach. A fully \textit{ab initio}  approach based on the non-equilibrium Green's function formalism~\cite{Pedroza_2017_bias_dependent} presents a very high computational hurdle for the simulation of rare reactive events. Other recent approaches like potentiostats that can be coupled to \textit{ab initio} simulations~\cite{Deissenbeck_2021_canon_potstat}, or grand-canonical \gls*{scf} techniques~\cite{Hagiwara_Otani_2021} could provide a good compromise between efficiency and accuracy, but still present an elevated computational cost.  
Effecitve 2D-periodic models like the \gls*{esm}~\cite{Otani_Sugino_2006} have allowed the simulation of \textit{ab initio} molecular dynamics of water at charged metal interfaces~\cite{Sugino_2007_bias_water, Otani_2008_Pt_bias_water} more than 15 years ago.
With the \gls*{esm}, the water dissociation on a platinum surface under potential bias could be studied by \gls*{aimd}~\cite{Ikeshoji_Hamada_2011_water_Pt111_bias, Qian_Hamada_2013_water_Pt111_bias}, which presents a remarkable achievement.
However, only relatively short \gls*{md} trajectories (of a few picoseconds) could be simulated, which considerably limits the predictive power of such simulations given that the reaction is a rare event in this timescale.
Interestingly, it has been recently shown from classical \gls*{md} simulations, that results obtained by simulations with a constant applied potential between two electrodes in a 2D-periodic model were indistinguishable from those obtained from a 3D-periodic model with a large metallic slab in the center of the simulation cell, and an applied electric field~\cite{Dufils_2019_slab_vs_capacitor} -- a possibility that can also simplify \textit{ab initio} simulations of electrochemical interfaces. 

In this paper, we employ an applied electric-field setup to address the water dissociation reaction directly, and understand the role of different surface charges and nuclear quantum effects in modifying the reaction dynamics.
We study a model system of a water monomer adsorbed on a Pd(111) surface, modelled by density-functional theory.
Different charges on the surface of the metal are realized by the application of an electric field on a thick Pd(111) slab, which mimics a potential bias.
We report the changes in reaction barriers and rates and the effects of zero-point energy at different field strengths.
These results serve as ground-work for more complex system setups and simulation protocols. They also give valuable insights on how far \gls*{nqe} and electric fields can impact this reaction.

\section{Methods}

\begin{table*}[htbp]
    \centering
    \caption{Adsorption energy $E_{\rm{ads}}$, potential energy barrier $E_a$,  \gls*{zpe}-corrected barrier of dissociation and corresponding quasi-harmonic quantum TST dissociation rates $k_d$ of a water monomer on a Pd(111) surface, depending on the applied electric field.
    The electric field of 1~V/{\AA} corresponds to the surface charge of 0.0364 electron per Pd atom (8.734 $\rm{C/cm^2}$). The values are calculated with \gls*{pbe} + vdW$^{\rm{surf}}$, unless specified otherwise. }
    \label{tab_Eads_monomer} \label{tab_Eact_monomer}
    \begin{tabular}{r|r|r|r|r|l}
    Electric field (V/{\AA}) & $E_{\rm{ads}}$ (eV) & $E_a$ (eV) & \acrshort*{zpe}-corrected $E_a$ (eV) & $k_d^{\text{300 K}}$ (s$^{-1}$) & Note \\
    \hline
    +0.74    & 0.40  & 1.32   & 1.12 & 4.3 $\times$ 10$^{-5}$ \\
    +0.44    & 0.44  & 1.18   & 0.98 & 8.6 $\times$ 10$^{-3}$ \\
    +0.15    & 0.47  & 1.08   & 0.86 & 5.3 $\times$ 10$^{-1}$ \\
    +0.07    & 0.48  & 1.06   & ---  & --- \\
    \hline
    no field & 0.48  & 1.03   & 0.83 & 4.7  & this work         \\
             & 0.31  & ---    & ---  & --- & this work, no vdW \\
             & 0.33  & ---    & ---  & --- & \cite{Michaelides_2003_water_ads_metals}   PW91, no vdW \\
             & ---   & 1.12   & ---  & --- & \cite{Wang_Tao_Bu_2006_water_dissoc_ANEBA} PBE,  no vdW \\
             & 0.22  & 1.09   & 0.87 & --- & \cite{Cao_2006_water_dissoc_CINEB}         PW91, no vdW \\
             & 0.30  & 1.05   & ---  & --- & \cite{Phatak_2009_water_dissoc_CINEB}      PW91, no vdW \\
             & 0.31  & 1.15   & 0.96 & --- & \cite{Fajin_2010_water_dissoc_DIMER}       PW91, no vdW \\
             & \rev{0.24}  & --- & --- & --- & \cite{Carrasco_Michaelides_2013_role_of_vdW}   \rev{PBE, no vdW} \\
             & \rev{0.42}  & --- & --- & --- & \cite{Carrasco_Michaelides_2013_role_of_vdW}   \rev{revPBE + vdW-DF*} \\
             & \rev{0.46}  & --- & --- & --- & \cite{Carrasco_Michaelides_2013_role_of_vdW}   \rev{optPBE + vdW-DF*} \\
             & \rev{0.53}  & --- & --- & --- & \cite{Carrasco_Michaelides_2013_role_of_vdW}   \rev{optB88 + vdW-DF*} \\
    \hline
    --0.07   & 0.45  & 1.02   & ---  & --- \\
    --0.15   & 0.41  & 1.00   & 0.80 & 1.4 $\times$ 10$^{+1}$ \\
    --0.29   & 0.34  & 0.98   & 0.78 & 1.9 $\times$ 10$^{+1}$ \\
    --0.44   & 0.31  & 1.02   & 0.85 & 2.4 $\times$ 10$^{-1}$ \\
    --0.74   & 0.30  & 1.01   & 0.90 & 1.1 $\times$ 10$^{-1}$ \\
    --1.00   & 0.30  & ---    & ---  & ---\\
    \hline
    \end{tabular} \\
    \rev{*see ref.~\cite{Klimes_2010_vdW_DF_modifications} for the details about the vdW-DF with different functionals.}
\end{table*}

We simulated a Pd(111) surface by a slab of 7 atomic layers in a periodic cell including a large vacuum layer of 64~{\AA}, using the FHI-aims code~\cite{FHIaims}. 
A dipole correction~\cite{Neugebauer_Scheffler_1992} was applied to compensate for spurious interactions between periodic images.
We employed the PBE exchange-correlation functional~\cite{PBE} augmented by 
dispersion interactions as described by the screened vdW$^{\rm{surf}}$ model~\cite{Ruiz2012}, with the coefficients for Pd taken from Ref.~\cite{Ruiz2016}.
\textit{Light} default settings of FHI-aims were used for basis sets, and modified \textit{light} settings with double radial density were used for numerical real-space grids.
The lateral dimensions of the Pd slab were chosen as 4$\times$4 unit cells, simulated with a 3$\times$3 k-points mesh. We placed the metallic slab at the center of the unit cell.

\begin{figure}[ht]
    \begin{center}
        \includegraphics[width=\columnwidth]{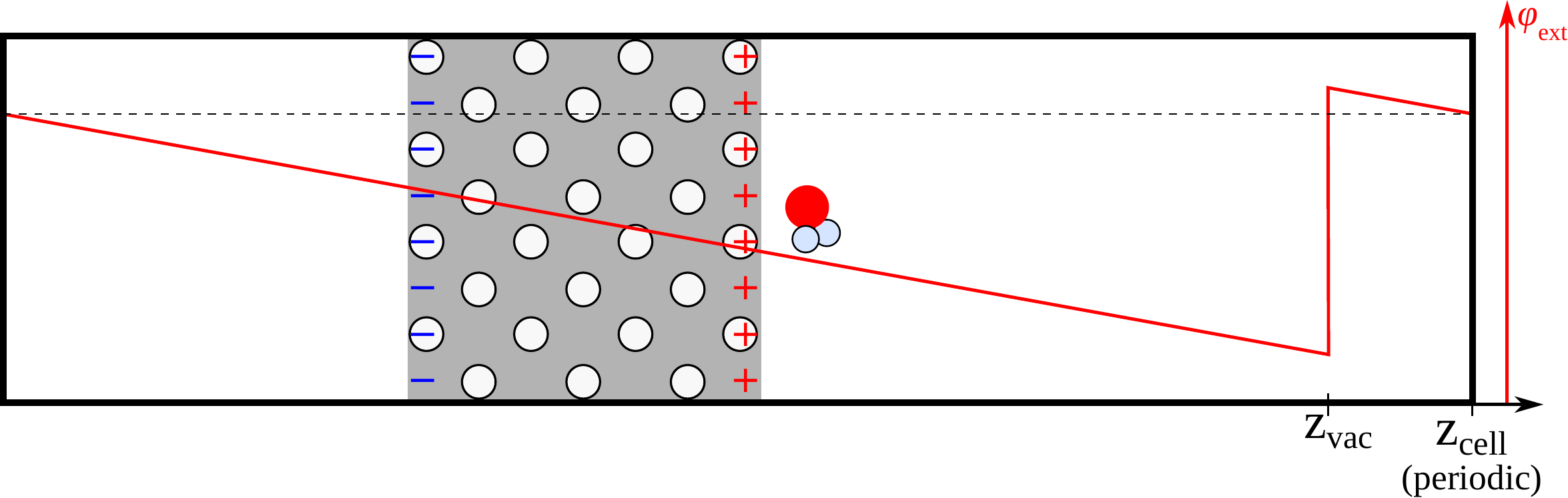}
        \caption{Schematic of simulation setup. A periodic cell is permeated by an external electrostatic potential $\varphi_{\rm{ext}}$ with a discontinuity in vacuum far away from the slab.}
        \label{fig_scheme}
    \end{center}
\end{figure}

The potential bias is mimicked by an external electric field permeating the system, which is implemented as a saw-tooth potential with the potential step in the vacuum region (see Fig.\ref{fig_scheme}). 
The polarization induced by the electric field causes the surfaces of the slab to effectively become capacitor plates with opposite charges.
We assured that the slab is thick enough to screen the surfaces from each other, by analyzing the charge density distribution at self consistency, under different field strengths.
Of course, the surfaces remain coupled by the charge conservation constraint.
We observed a linear dependence of the surface charge on the applied electric field in a range from --10 to +10~V/{\AA}. An increase of the field by 1~V/{\AA} induced a surface charge of 0.0364 electron per Pd atom, which is equivalent to 8.734 $\rm{C/cm^2}$.
The surfaces at zero electric field are slightly negatively charged, which offsets the linear dependence by --0.012~e per atom.
We will call a field ``negative'' or ``positive'' according on the charge that it induces to the surface of interest.

\rev{The adsorption geometries of a water molecule on a Pd(111) surface were obtained by relaxation from the high-symmetry sites of the surface: atop, bridge and fcc/hcp hollow sites. If more than one geometry was found after the relaxation, the lowest-energy structure was chosen.}

For the phonon calculations, a modified version~\cite{fidanyan_phonopy} of the Phonopy package~\cite{phonopy} was used, and either only the water molecule, or the molecule and the first layer of the surface were included to build the Hessian.
The deeper layers of the surface were included as a rigid environment.
By doing so, we lose the coupling Hessian elements between the atoms of the molecule and the surface beyond the first layer, which can affect low-frequency molecule-surface phonons.
Since the molecule is small and light, we assume that long-wave collective vibrations of a surface do not play a significant role in water reactions, 
because the chemically relevant frequencies of the H-O-H bending and O-H stretching are not substantially affected.

The reaction paths were obtained using the \gls*{cineb} method~\cite{CI_NEB_2000}.
We used a new implementation of \gls*{cineb} in the i-PI~\cite{iPIv2} code, in which several instances of the FHI-aims code could connect simultaneously to i-PI and efficiently calculate the forces on each replica in parallel.
This implementation is described in detail in Ref.~\cite{KarenThesis2022} and is available through the main repository of the code. 
We used a NEB path with 9 intermediate replicas with the spring strength of 20-40~eV/\r{A}. We used the FIRE~\cite{Bitzek_2006_FIRE} algorithm for the optimization. The tolerance for forces was set to 0.05~eV/\r{A} for the optimisation of the NEB path and to 0.01~eV/\r{A} for the climbing-image optimisation.

\section{Adsorption of water on a charged Pd(111) surface} \label{sec_water_adsorption}

\begin{figure}[ht]
    \begin{center}
        \includegraphics[width=0.9\columnwidth]{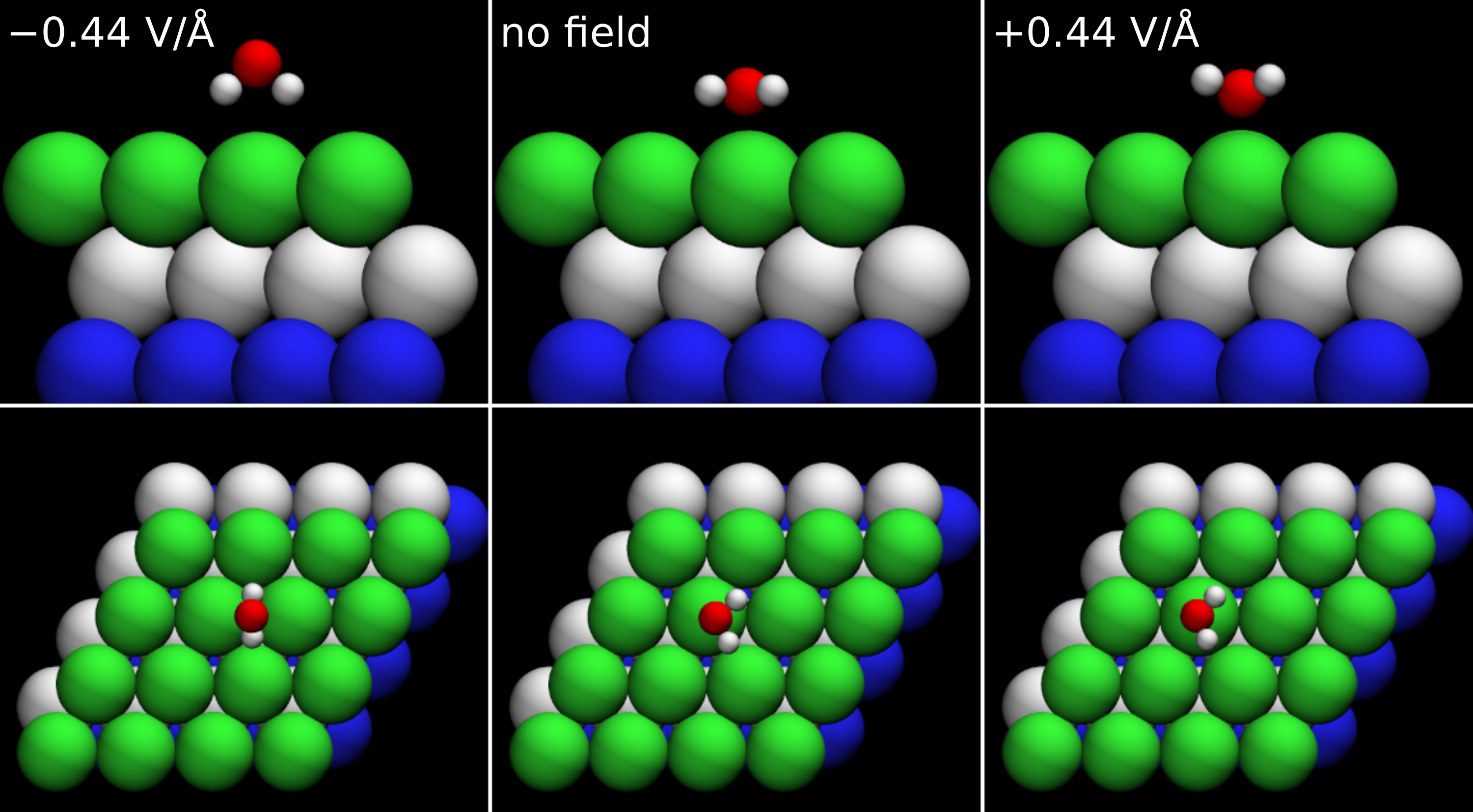}
        \caption{Adsorption geometries of a single water molecule on the Pd(111) surface under --0.44~V/{\AA} (left), no field (center) and +0.44~V/{\AA} (right).
        Large green, white and blue spheres denote the 1st, 2nd and 3rd layers of the Pd(111) surface, respectively.}
        \label{fig_intacts}
    \end{center}
\end{figure}

We have found the preferred adsorption geometries of a single water molecule at the Pd(111) surface for different electric field strengths and report them in  Fig.~\ref{fig_intacts}.

\begin{figure}[ht]
    \begin{center}
        \includegraphics[width=1.0\columnwidth]{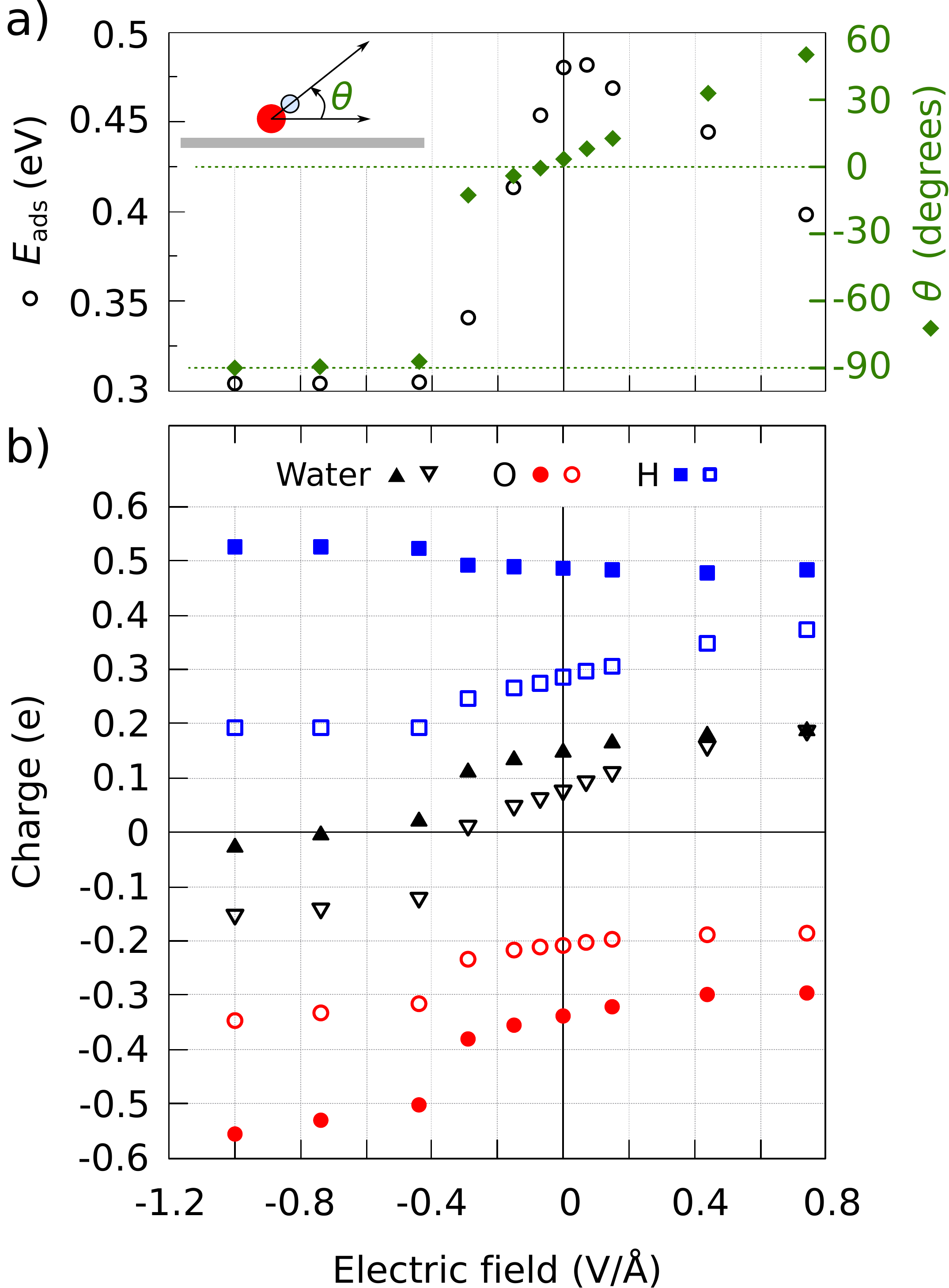}
        \caption{\rev{a) Empty black circles, left-hand axis: adsorption energy of a single water molecule on the Pd(111) surface as a function of the applied electric field.
        The values are calculated with \gls*{pbe} + vdW$^{\rm{surf}}$.
        Green diamonds, right-hand axis: angle $\theta$ between the water molecule and the Pd surface (see the inset, the hydrogen atoms are equivalent).}
        b) Charge induced on the adsorbed water molecule depending on the applied electric field, measured by the Mulliken (filled points) and Hirshfeld (empty points) analysis.
        The black triangles show the total charge on the molecule, and the blue squares (red circles) show H (O) contributions to it.
        }
        \label{fig_charges_and_Eads}
    \end{center}    
\end{figure}
We found that the water molecule prefers a flat orientation if no field is applied, which agrees with previous works~\cite{Wang_Tao_Bu_2006_water_dissoc_ANEBA, Cao_2006_water_dissoc_CINEB, Phatak_2009_water_dissoc_CINEB}.
In a negative electric field, the flat orientation remains up to --0.37~V/{\AA}, after which the molecule ``stands up'' pointing both hydrogen atoms towards the surface.
In a positive field, the molecule deviates gradually from the flat geometry, and hydrogen atoms turn more and more towards the vacuum. These observations are also in qualitative agreement with earlier work, for example regarding a water monomer on Au(111) treated with non-equilibrium Green's functions~\cite{Pedroza_2017_bias_dependent}.
We did not observe any abrupt changes in the range between --0.37 and +1~V/{\AA}, \rev{as evidenced by the angles between the plane defined by the water molecule and the surface plane, reported in Fig.~\ref{fig_charges_and_Eads}a.}

We calculated the adsorption energies of water $E_{\rm{ads}}$ in the presence of an electric field. The references to calculate $E_{\rm{ads}}$ at each field strength were taken as the isolated subsystems optimized in the presence of the respective electric field embedding.
\rev{In addition, we report the values of the adsorption energy related to the zero-field reference in the SI.}
The adsorption energies are summarized in Fig.~\ref{fig_charges_and_Eads}a and Table~\ref{tab_Eads_monomer}.
The highest adsorption energy is observed when no electric field or a slight positive field of 0.07~V/{\AA} is applied.
The value obtained without an electric field is 0.48~eV, which is larger than the values of 0.22-0.31~eV reported in previous studies~\cite{Cao_2006_water_dissoc_CINEB, Phatak_2009_water_dissoc_CINEB, Fajin_2010_water_dissoc_DIMER}. The reason for this discrepancy is that no \gls*{vdw} interactions were included in these previous studies, as we show in Table~\ref{tab_Eads_monomer}.
\rev{Inclusion of the vdW interactions in Ref.~\cite{Carrasco_Michaelides_2013_role_of_vdW} result in adsorption energies very similar to the ones we obtain.} 
When increasing the field towards positive values, $E_{\rm{ads}}$ gradually decreases (i.e., the adsorbed system becomes less stable).
When a negative field is applied, $E_{\rm{ads}}$ also gradually decreases down to a field of --0.44~V/{\AA}, after which it remains approximately constant.
This change of the trend corresponds to the flip of the preferred adsorption orientation discussed above.

\begin{figure}[ht]
    \begin{center}
        \includegraphics[width=1\columnwidth]{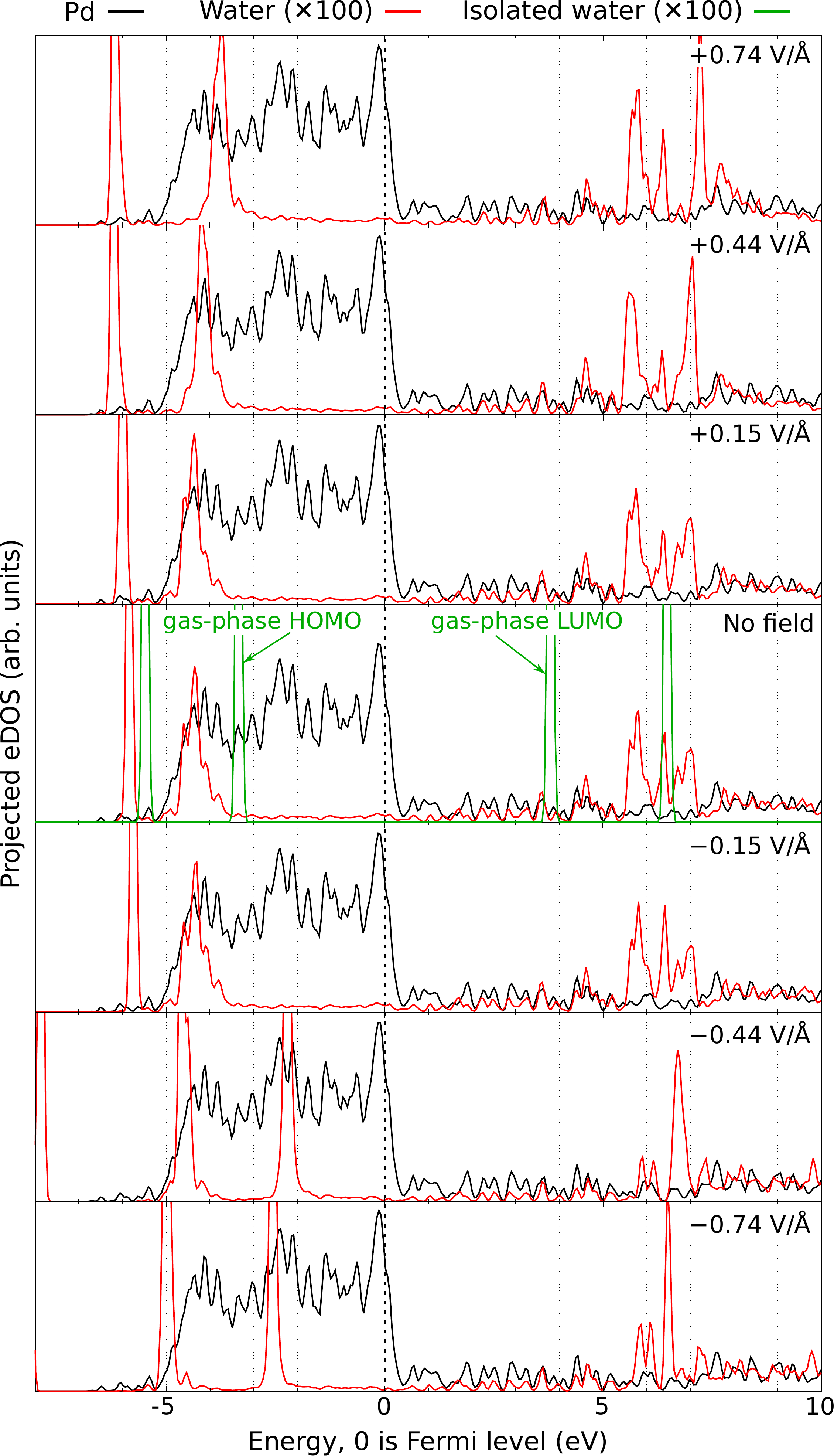}
        \caption{Electronic density of states projected on the atomic species at different values of an external electric field.
            The \rev{black} lines denote Pd and the \rev{red} lines the sum of O and H contributions.
            Green curves on the zero-field plot show the \gls*{dos} of an isolated water molecule, aligned so that the 1s orbital of its oxygen matches \rev{the 1s orbital of the adsorbed water molecule, after which all values are shifted such that zero corresponds to the Fermi level of the calculations including Pd(111).}}
        \label{fig_edos}
    \end{center}
\end{figure}

We investigated how the character of the bond between the molecule and the surface changes with an electric field by projecting the total electronic \gls*{dos} on the atoms. 
We show such projected \gls*{dos} in Fig.~\ref{fig_edos}.
By comparing the positions of the molecular orbitals in the gas-phase to those on the surface, we observe in Fig.~\ref{fig_edos} that what used to be the HOMO level of the water molecule in isolation becomes substantially broader for field values ranging from +0.74 to --0.15 V/\AA, which points toward a hybridization with the Pd $d$-bands. For more negative fields, the hybridization is drastically reduced, which is consistent with the ``standing'' orientation of the water molecule and the character of the HOMO level of water (centered on oxygen). 
This is also reflected on the Hirshfeld and Mulliken analysis of the charge accummulated on the molecule, shown in Fig.~\ref{fig_charges_and_Eads}b.
On a positively charged surface, Hirshfeld and Mulliken analysis agree for the total charge, even if they predict different values for the individual atoms.
The molecule donates roughly $0.2$~e to the surface, indicating orbital hybridization and electron transfer from the molecule to the surface.
At an electric field below --0.44~V/{\AA}, when the molecule ``stands'', the Mulliken analysis shows, in contrast, a negligibly small charge, while the Hirshfeld method shows about --$0.15$~e.
We interpret the discrepancy between the Hirshfeld and Mulliken charges as a specific feature of the Hirshfeld definition of charge, which assigns local changes of the electron density to atoms proportionally to their free-atom electron density at the considered point.
Therefore, a change of the ``tail'' of surface density can be assigned to the molecule even if there is no hybridization of the orbitals of the surface and the molecule.

\section{Reaction paths in an electric field} \label{sec_neb_water}

\begin{figure*}[ht]
   \includegraphics[width=\textwidth]{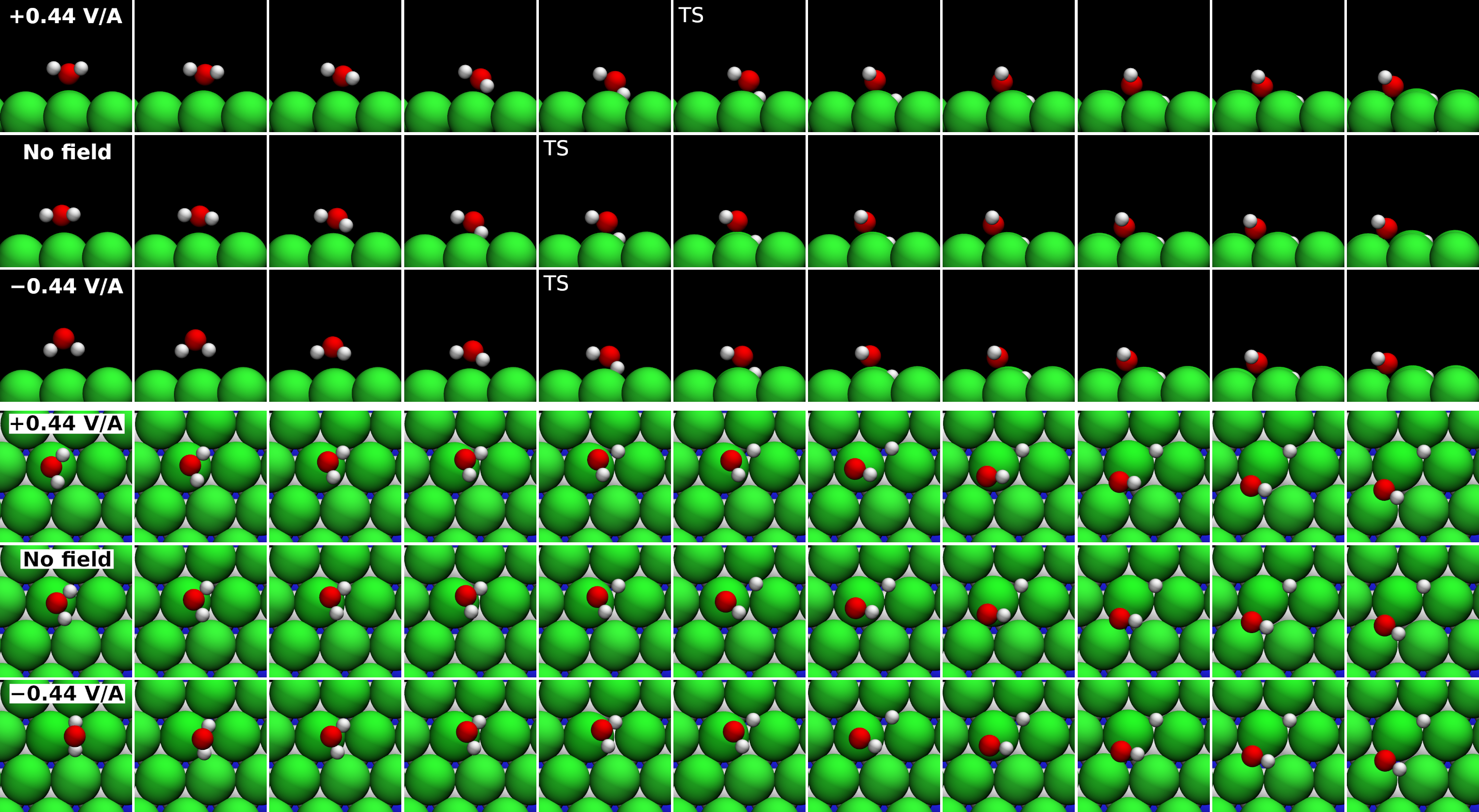}
        \caption{\rev{Individual \gls*{neb} node geometries of the reaction path found by the CI-NEB algorithm, side and top view.
        Large green spheres denote the 1st layer of the Pd(111) surface, and the white and blue spheres of the 2nd and 3rd Pd layers are visible in the HCP- and FCC-hollow sites of the surface, respectively.
        }}
        \label{fig_frames_new_layout}
\end{figure*}

\begin{figure}[ht]
    \begin{center}
        \includegraphics[width=1\columnwidth]{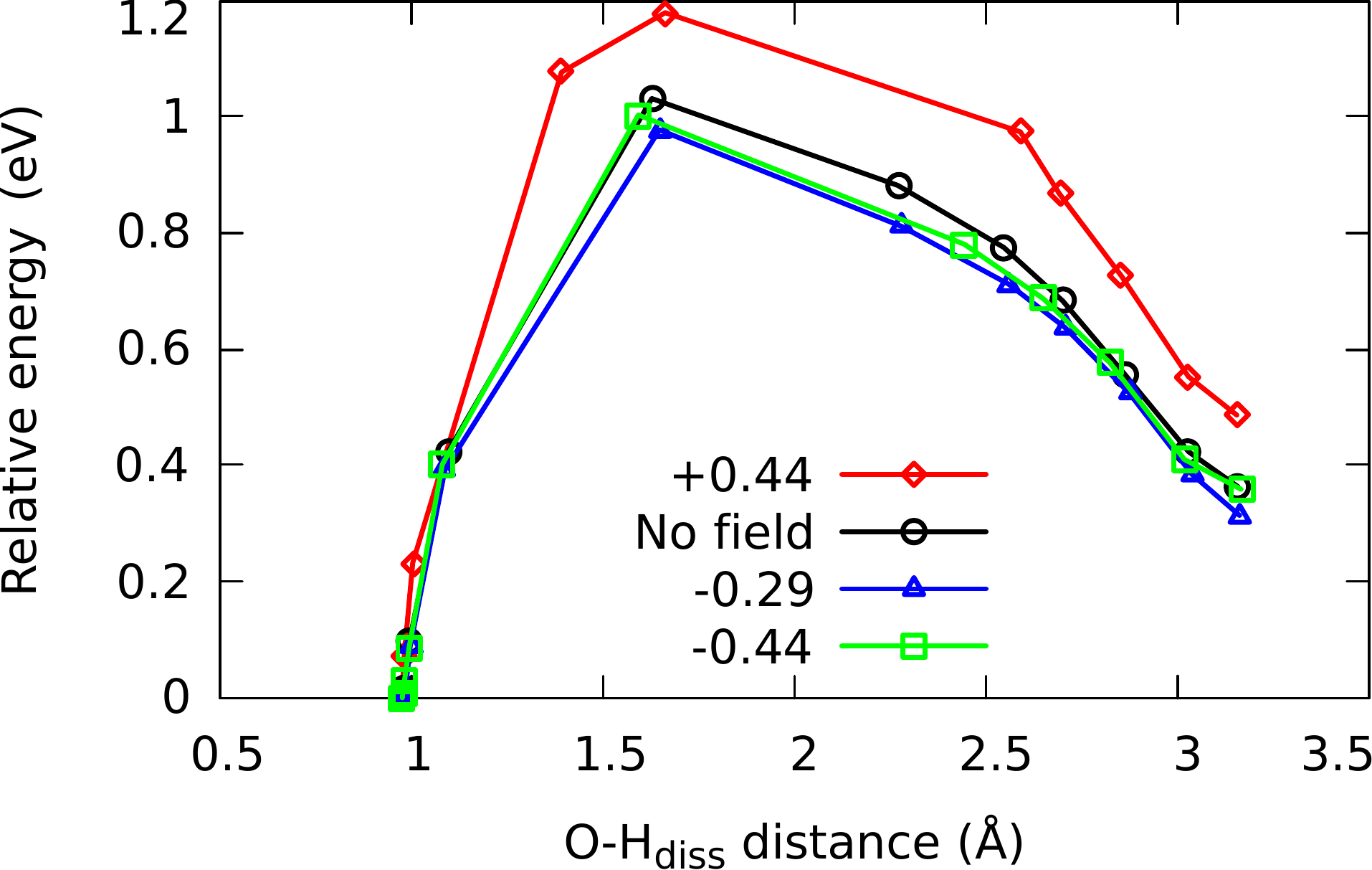}
        \caption{\rev{Potential energy along the \gls*{mep} found by the \gls*{cineb} algorithm.
        The points represent the energy of the NEB nodes relative to the reactant, and the $x$ axis shows the corresponding distance between the oxygen atom and the dissociated hydrogen.
        The barriers are shown for the electric field of +0.44~V/\r{A} (red diamonds), no field (black circles), --0.29~V/\r{A} (blue triangles) and --0.44~V/\r{A} (green squares).}
        }
        \label{fig_barriers_vs_oh_dist}
    \end{center}
\end{figure}

\begin{figure}[ht]
    \begin{center}
        \includegraphics[width=1\columnwidth]{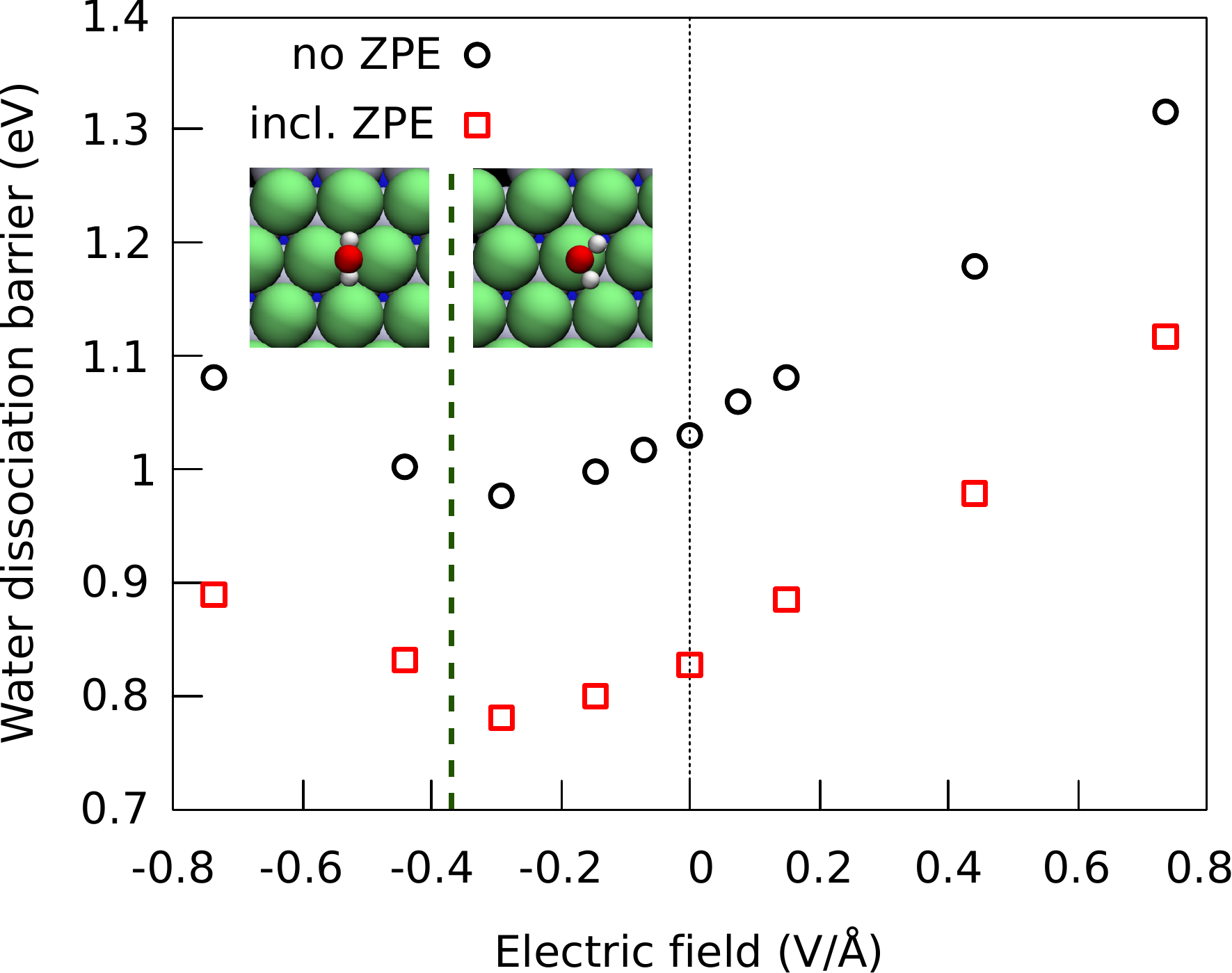}
        \caption{The energy barrier of water splitting calculated by \gls*{cineb} algorithm.
        Black circles show the potential energy barrier, and red squares show the ZPE-corrected barrier, calculated in the harmonic approximation.
        The dashed green line marks the border between two orientations of the reactant state\rev{, which are shown schematically in the insets}.
        }
        \label{fig_barrier_vs_voltage}
    \end{center}
\end{figure}

The \gls*{neb} paths are shown in figure~\ref{fig_frames_new_layout}, and corresponding energy barriers in figure~\ref{fig_barriers_vs_oh_dist}. 
We verified the transition states by a normal mode analysis.
In all cases, the transition state has exactly one imaginary-frequency mode, which corresponds to a direction of barrier crossing.
The geometry of the transition state has little dependence on the surface charge, especially if compared to the reactant state.
The activation energy $E_a$ for each bias is summarized in Fig.~\ref{fig_barrier_vs_voltage} and in Table~\ref{tab_Eact_monomer}.


For zero bias, our $E_a$ value lies below (by 20-120~meV) previously reported calculations~\cite{Wang_Tao_Bu_2006_water_dissoc_ANEBA, Cao_2006_water_dissoc_CINEB, Phatak_2009_water_dissoc_CINEB, Fajin_2010_water_dissoc_DIMER}, which also did not include dispersion interactions.
Many previous work employed the PW91 \gls*{xc} potential, but it was shown on various metals that the water dissociation barrier calculated by PW91 and \gls*{pbe} are almost identical, e.g. see the SI of Ref.~\cite{Fajin_2010_water_dissoc_DIMER}.
Thus, the discrepancy with literature is consistent with the observation of Litman \textit{et al.} that inclusion of \gls*{vdw} interactions decreases the barrier of dissociation of water on metals~\cite{Litman_2018_SLRPC}.
The lowest barrier is observed at --0.29~V/{\AA}, directly before the water molecule changes its adsorption orientation.  Positive electric fields only increase $E_a$. \rev{It is reasonable to expect that the availability of negative charges on the surface stabilizes the adsorbed OH$^-$ radical and favors the reaction, but there is more to it.} Analysing the \gls*{mep}, we speculate that positive field strengths stabilize the O-Pd bond at the reactant, as evidenced by the charge-transfer behaviour. This strengthening makes it harder for water to move towards the point where the TS is. Negative field strengths destabilize the bond, matching, in fact, the strength of the O-Pd bond at around the flipping point. Beyond the flipping point, the molecule is  physisorbed, but has to flip back in order to reach the transition state (see \gls*{mep}s in \rev{Fig.~\ref{fig_frames_new_layout}}), which again costs energy.

As discussed previously in the literature, the main nuclear quantum effect that affects water dissociation on metallic surfaces \rev{around room temperature} is zero-point-energy, with tunneling playing a \rev{minor} role~\cite{Litman_2018_SLRPC}. We confirm this observation here by an analysis of the frequency of the TS mode at all electric field strengths, reported in~\rev{Table~\ref{tab_crossover_temp}}. We estimate tunneling crossover temperatures $T_c$ to be  below \rev{185} K for all field strengths, \rev{obtaining a very similar value as the one reported in Ref.~\cite{Cao_2006_water_dissoc_CINEB} for this reaction without the application of a field. We remark that the highest $T_c$ is observed for --0.74 V/\AA, and that the field strengths reported here can change $T_c$ by 80 K. This observation shows that the application of the field and the consequent negative charging of the surface can considerably increase the curvature of the barrier (assuming its shape around the TS is reasonably approximated by an inverted parabola) and increase the importance of tunneling at higher temperatures in this reaction.} 

\rev{\begin{table}
    \caption{Imaginary frequency of the unstable mode at the transition state ($\omega^{\rm{TS}}$) and the tunneling crossover temperature ($T_c$) for water dissociation on Pd(111) surface in different electric fields.}
    \centering
    \rev{\begin{tabular}{c|c|c}
       Electric field (V/{\AA}) & $\omega^{\rm{TS}}$ ($i*\rm{cm}^{-1}$) & $T_c$ (K)  \\ \hline \hline
        +0.74 & 451.7&  103.4 \\
        +0.44 & 564.4 &  129.2 \\
        +0.15 & 621.7 &  142.4\\
        0     & 645.4 &  147.8 \\
       --0.15 & 673.1 &  154.1 \\
       --0.29 & 640.5 &  146.7 \\
       --0.44 & 729.5 &  167.0 \\
       --0.74 & 800.5 &  183.3 \\
    \hline
    \end{tabular}}
    \label{tab_crossover_temp}
\end{table}}

We \rev{calculate} the harmonic \gls*{zpe} contributions to reactant and transition states, and summarize the changes of the dissociation barrier in Fig.~\ref{fig_barrier_vs_voltage} and in Table~\ref{tab_Eact_monomer}. We note that a low-frequency vibrational mode of the reactant, corresponding to the rotation of water hydrogens around the oxygen, parallel to the surface plane, becomes progressively lower in frequency and even slightly negative (\rev{$\approx -40$} cm$^{-1}$) towards higher positive field values. We have tightened optimization thresholds and increased numerical and basis set accuracies to confirm this observation. First, we can conclude that this model system is indeed slightly unstable above +0.44 V/\AA~ within the harmonic approximation, and the molecule could be freely rotating parallel to the surface plane. Second, these modes do not play a significant role on the water dissociation path and, because this is a feature of the model that would disappear if water was embedded in an H-bonded network, for the ZPE analysis we consistently ignored this mode in all calculations. We should add that because the mode is so low in energy (always below 30 cm$^{-1}$), its contribution would always be less than 6 meV to the total ZPE.

\begin{figure}[ht]
    \begin{center}
        \includegraphics[width=1\columnwidth]{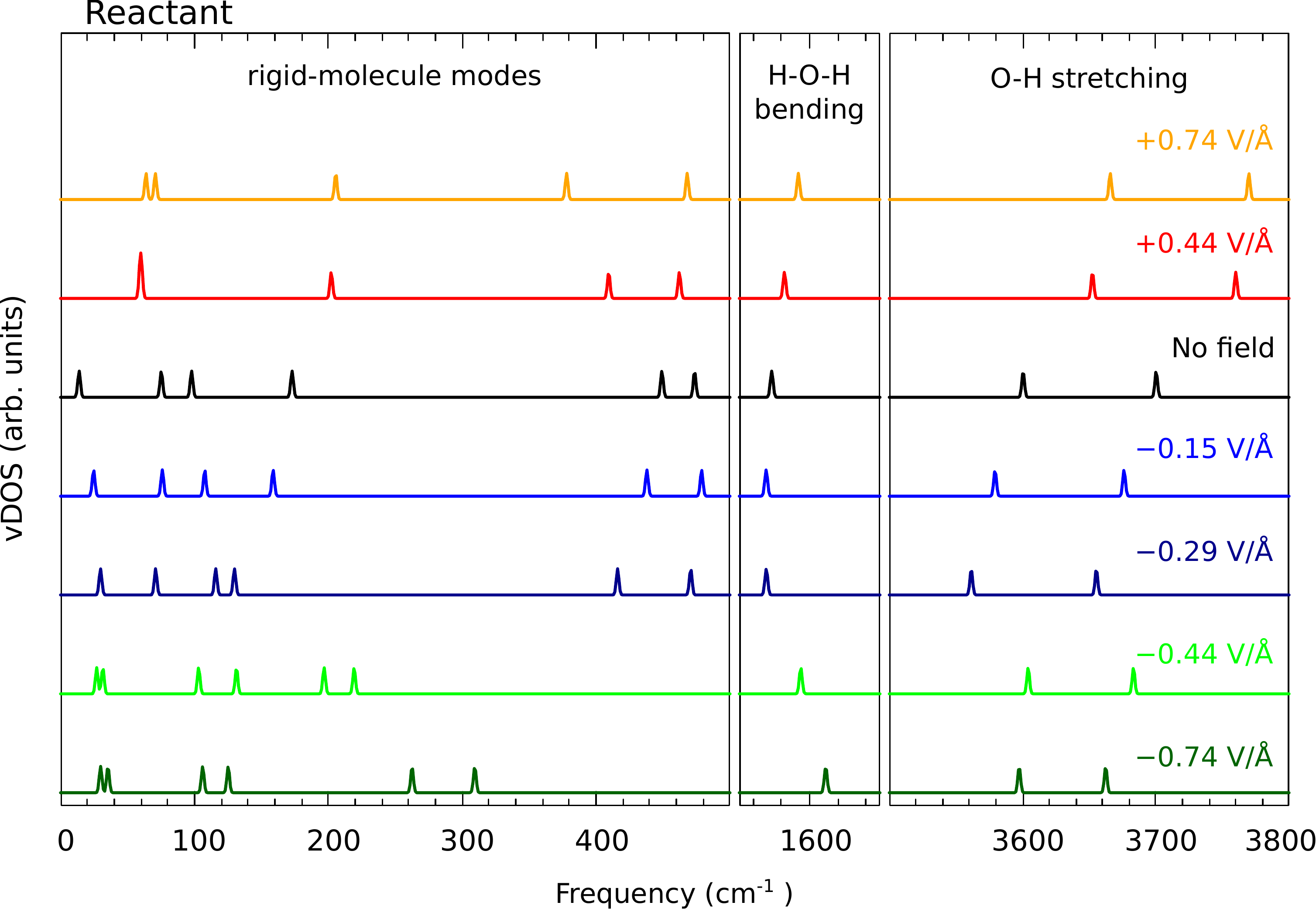}

        \vspace{10 pt}

        \includegraphics[width=1\columnwidth]{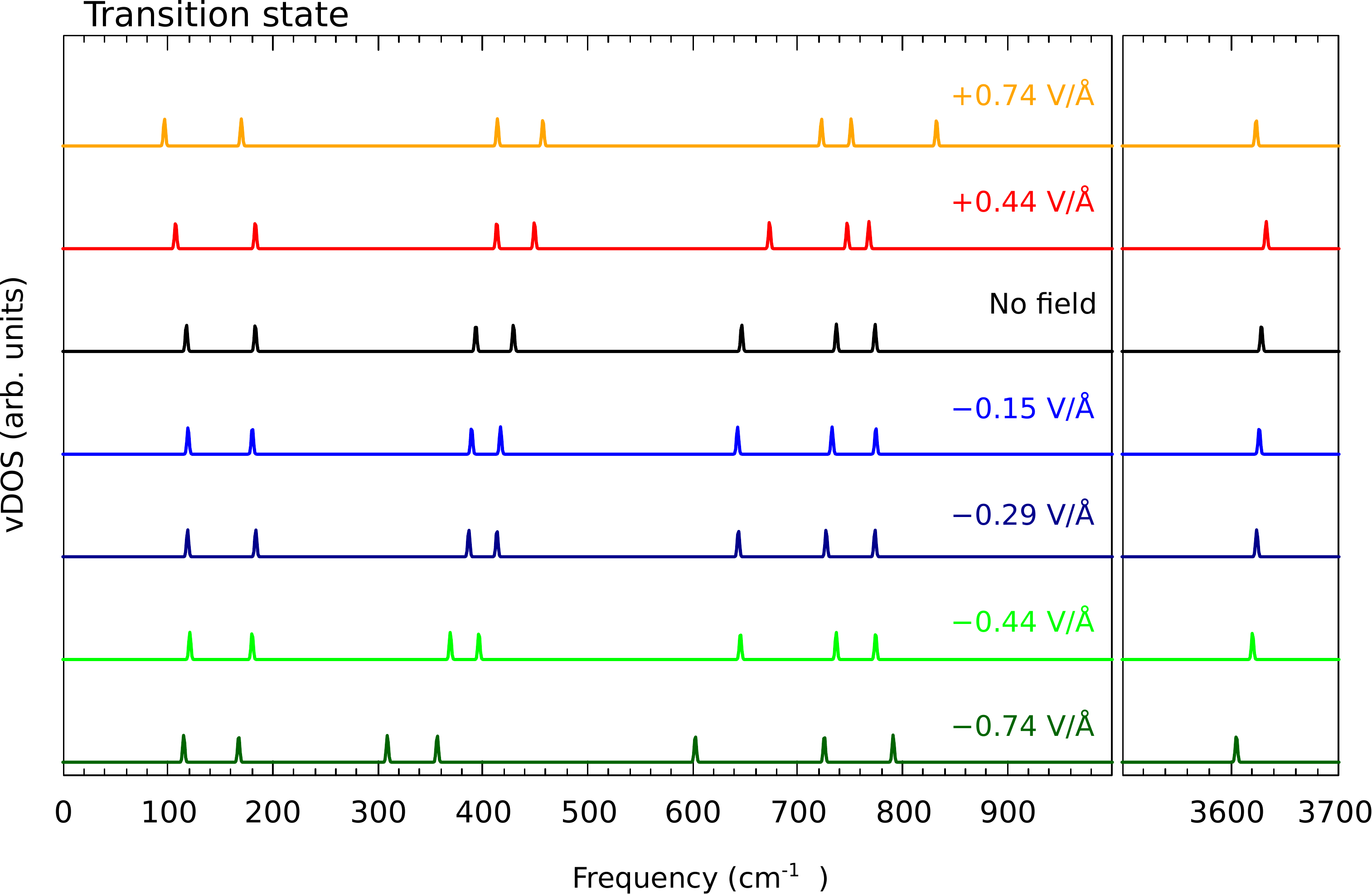}
        \caption{Vibrational density of states of the reactant state (top) and of the transition state (bottom) of the water splitting reaction on a Pd(111) surface at different electric field values. The lowest-frequency mode of the reactant state becomes unstable at +0.44 and +0.74~V/\r{A}, therefore it is not shown.}
        \label{fig_vDOS}
    \end{center}
\end{figure}

\begin{figure}[ht]
    \begin{center}
        \includegraphics[width=0.6\columnwidth]{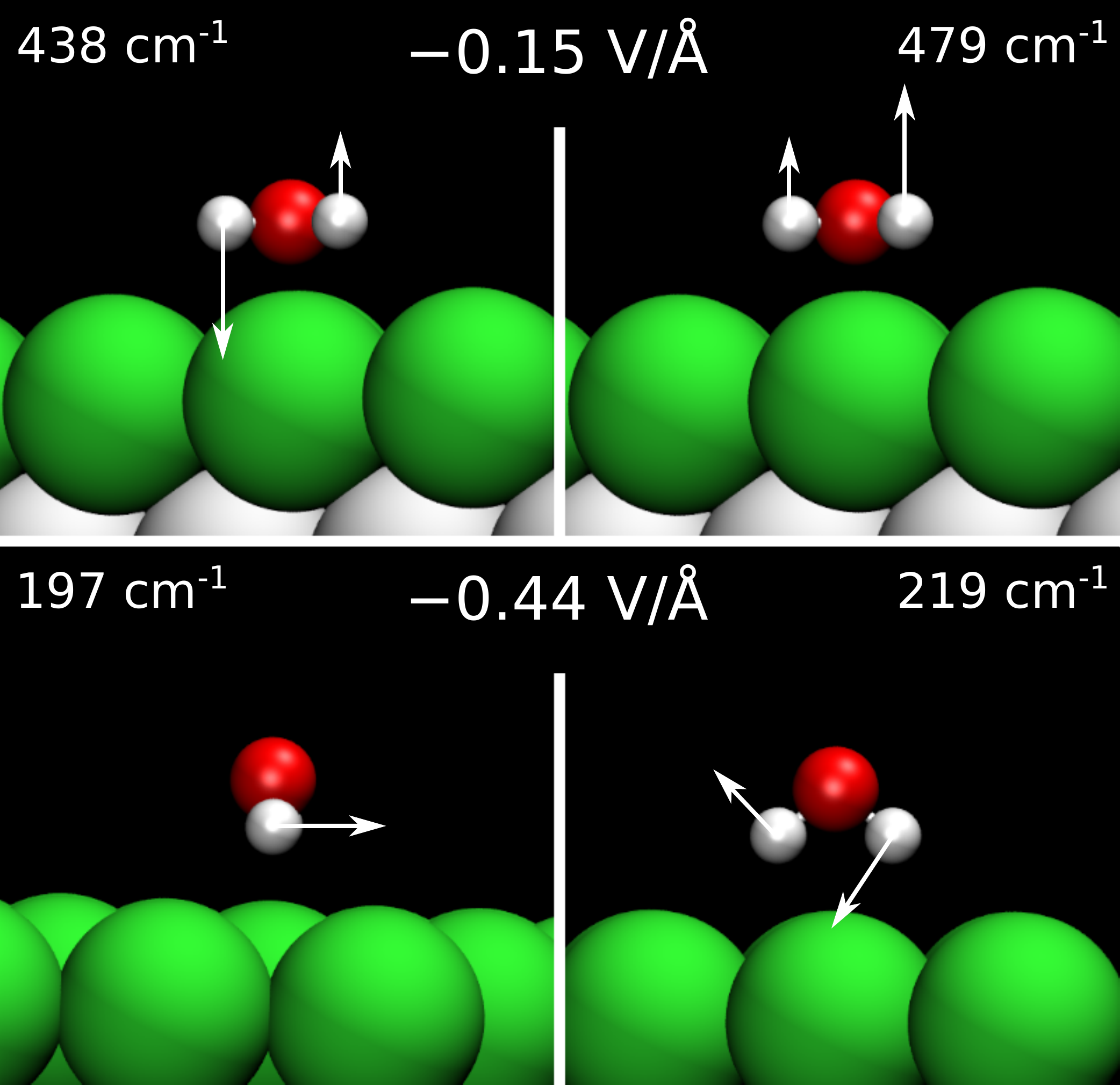}
        \caption{Hindered rotation modes of a water molecule on a Pd(111) surface at --0.15~V/{\AA} (top) and --0.44~V/{\AA} (bottom).
       }
        \label{fig_rigid-vibs}
    \end{center}
\end{figure}

The \gls*{zpe} contribution decreases the barrier by roughly 0.19-0.20~eV at all points except --0.44~V/{\AA}, where the decrease is 0.17~eV,~\rev{as shown in Fig~\ref{fig_barrier_vs_voltage}}. This is a considerable contribution in this case, representing $\approx 20$\% of the barrier height.
However, such a constant effect is somewhat surprising, given that the geometry of the reactant state changes considerably.
In order to resolve the effects of \gls*{zpe}, we analyzed the \gls*{vdos} spectra of the reactant and transition states, which are shown in Fig.~\ref{fig_vDOS}, top and bottom, respectively.
The most pronounced difference appears in the two hindered rotation modes of the reactant, which shift sharply from the range of 400-500~cm$^{-1}$ down to 200-320~cm$^{-1}$ when the field reaches --0.44~V/{\AA} (the flipping point).
These modes are shown in Fig.~\ref{fig_rigid-vibs}. It is important to note that hindered rotation modes are expected to be strongly anharmonic and thermally populated, as quantified in Ref.~\cite{RossiJCPPerspective2021}, therefore the harmonic analysis presented here gives only an estimate of the effect.

Despite multiple shifts in the low-frequency part of the spectrum, the  \gls*{zpe} is dominated by the high-frequency modes, namely the H-O-H bending mode and one of the two O-H stretching modes, which exist in the reactant state and disappear at the transition state. This is the main source of difference between the reactant and transition state contributions. We present a detailed account of the \rev{contribution} of particular modes on the total ZPE in the SI. Even though the \rev{high-frequency} modes also present a strong variation in the reactant state under different applied electric fields, the ZPE contributions are compensated between product and reactant states, as an analysis of the cumulative ZPE contribution shows (see SI), resulting in the essentially constant ZPE correction to the reaction barrier.

In order to estimate how much the reaction rates can be changed due to the electric field effect, we report reaction rates as calculated from the quantum quasi-harmonic approximation (Eq. 13 in Ref.~\cite{Litman_2018_SLRPC}). We report these rates at a temperature of 300 K in Table~\ref{tab_Eact_monomer}. This is an artificial temperature because the single molecule would have a considerable desorption probability at this temperature and would surely be very mobile. However, a single adsorbed water molecule is already a model system in this context. Therefore, these numbers serve only to build a more quantitative intuition. As expected, the suppression of the reaction rate at positive field values (up to 5 orders of magnitude) is much more significant than its increase at negative field values (only up to a factor of 4). We predict that the flipping point of the first water layer will also limit the increase of this reaction rate when water is at the liquid state. Because in the liquid, for the first water layer to flip, hydrogen bonds would need to be broken, the flipping point should happen at considerably larger \rev{effective} field strengths, \rev{even if it is difficult to give a quantitative estimate due to entropic contributions in the liquid. We know from our calculations that the charge on the surface changes by 0.036 e/(V/\AA) per surface Pd atom, which puts the flipping point ($\approx -0.4$ V/\AA) at an excess surface charge of around $-0.014$ e/Pd. This excess surface charge can be connected to other studies in the liquid, as for example Ref.~\cite{OtaniJPSJ2008}, in which larger surface excess charges ($\approx -0.08$ e/Pt) are reported for a more pronounced structuring of the first water layer on a Pt surface.} 

\section{Conclusions}

We presented an \textit{ab initio} study of water dissociation on the Pd(111) surface using a slab model embedded in an electric field in a 3D periodic setup. The electric field effectively induces opposite charges on each surface of the thick metallic slab, and mimics the effect of an applied potential bias at the metallic electrode surface. 

Analysing the reaction paths at different field strengths, we concluded that the main feature that affects the reaction barrier is the geometry of the reactant state, which is strongly influenced by the applied field. Between field strengths of --0.3 and --0.4 V/\AA, the reactant state abruptly changes its geometry, with the water molecule pointing the hydrogen atoms to the surface.
We demonstrated that the lowest barrier (and consequently highest reaction rate) for the monomer dissociation lies at the electric field strength at which the molecule has these two adsorption geometries equally preferred, i.e., at the flipping point. We rationalized that at this point, the water-surface bond is optimally weakened, making it easier for the system to reach its transition state. 

\rev{Further, we showed that the application of negative electric fields, which in the convention used in this paper causes negative charges to be accumulated at the surface, increases the tunneling crossover temperature significantly, albeit remaining below 200 K for this reaction and the field strengths studied in this work.} \gls*{nqe} \rev{were explicitly evaluated} limited to \gls*{zpe} in the harmonic approximation, \rev{demonstrating} that \rev{the reaction barrier is reduced} by roughly 0.2~eV at all field strengths.
The reason for such a weak dependence on the field strength is the dominant role of bond stretching modes on the ZPE of the reaction and the mutual cancellation of contributions of different vibrational modes between reactant and transition states. \rev{The observation that the \gls*{zpe} contributions are almost constant throughout a large variety of field strengths could simplify more involved simulation protocols involving \gls*{md} simulations, where one could estimate \gls*{nqe} corrections at a single field strength and propagate that to others -- even if with care, because the effect of \gls*{nqe} may not be uniform across field strengths for different liquid water properties~\cite{CassoneSaija2022review}.}

Although we studied a highly idealized system, we expect these findings to be relevant for more complex scenarios. Chemistry under applied electric fields is a booming field in its own right, and extending some of the protocols presented in this paper for methods that can capture nuclear tunneling could open exciting new avenues for situations where such effects play a role~\cite{Kozuch2020}.
In electrochemistry, the change in orientation of the first water layer under different potential biases has been observed experimentally~\cite{Toney_1994_voltage_dep_ordering_expt} and the role of these effects in the chemistry that happens at these interfaces is still not fully understood. 
\rev{We believe that further studies in liquid water, taking into account the insights obtained here, namely (i) that the reaction barrier is lowest at the flipping point of the reactant, (ii) that nuclear tunneling contributions are more significant on negatively charged surfaces and (iii) that ZPE has a large impact on the dissociation rate, should lead to a better understanding of the factors determining the water splitting reaction on  metallic substrates.} Finally, we note that we constrained the total charge of the simulation cell to be zero, while a real electrode would allow charge fluctuations on a larger scale and, possibly, with a lower energy penalty --  a possibility that should be explored in future work.

\begin{acknowledgements}
We acknowledge fruitful discussions with Alexandre Reily Rocha and Luana Sucupira Pedroza, who have immensely helped us navigate this subject. \rev{We further thank Yair Litman and Luana Sucupira Pedroza for reading a preliminary version of this manuscript and giving useful suggestions}. We would like to thank the Max Planck Computing and Data Center (MPCDF) for computing time. K.F. acknowledges support from the International Max Planck Research School for Ultrafast Imaging \& Structural Dynamics (IMPRS-UFAST). 
\end{acknowledgements}

\section*{Data Availability Statement}

Phonon calculations for the initial and transition states of the water splitting reaction, and the geometry relaxations of the adsorbed reactants and products for different values of an electric field are uploaded to the NOMAD repository. The dataset DOI is 10.17172/NOMAD/2022.09.15-1.

\bibliography{literature}

\end{document}


\title{Supplementary material: Ab initio study of water dissociation on a charged Pd(111) surface}

\author{Karen Fidanyan}
\affiliation{Max Planck Institute for the Structure and Dynamics of Matter, Luruper Chaussee 149, 22761 Hamburg, Germany}

\author{Guoyuan Liu}
\affiliation{Department of Materials Science and Engineering, École Polytechnique Fédérale de Lausanne, CH-1015 Lausanne, Switzerland}

\author{Mariana Rossi}
\email{mariana.rossi@mpsd.mpg.de}
\affiliation{Max Planck Institute for the Structure and Dynamics of Matter, Luruper Chaussee 149, 22761 Hamburg, Germany}

\maketitle

\section{Effect of the field on the adsorption energy}

\begin{figure}[h]
    \centering
    \includegraphics[width=0.75\textwidth]{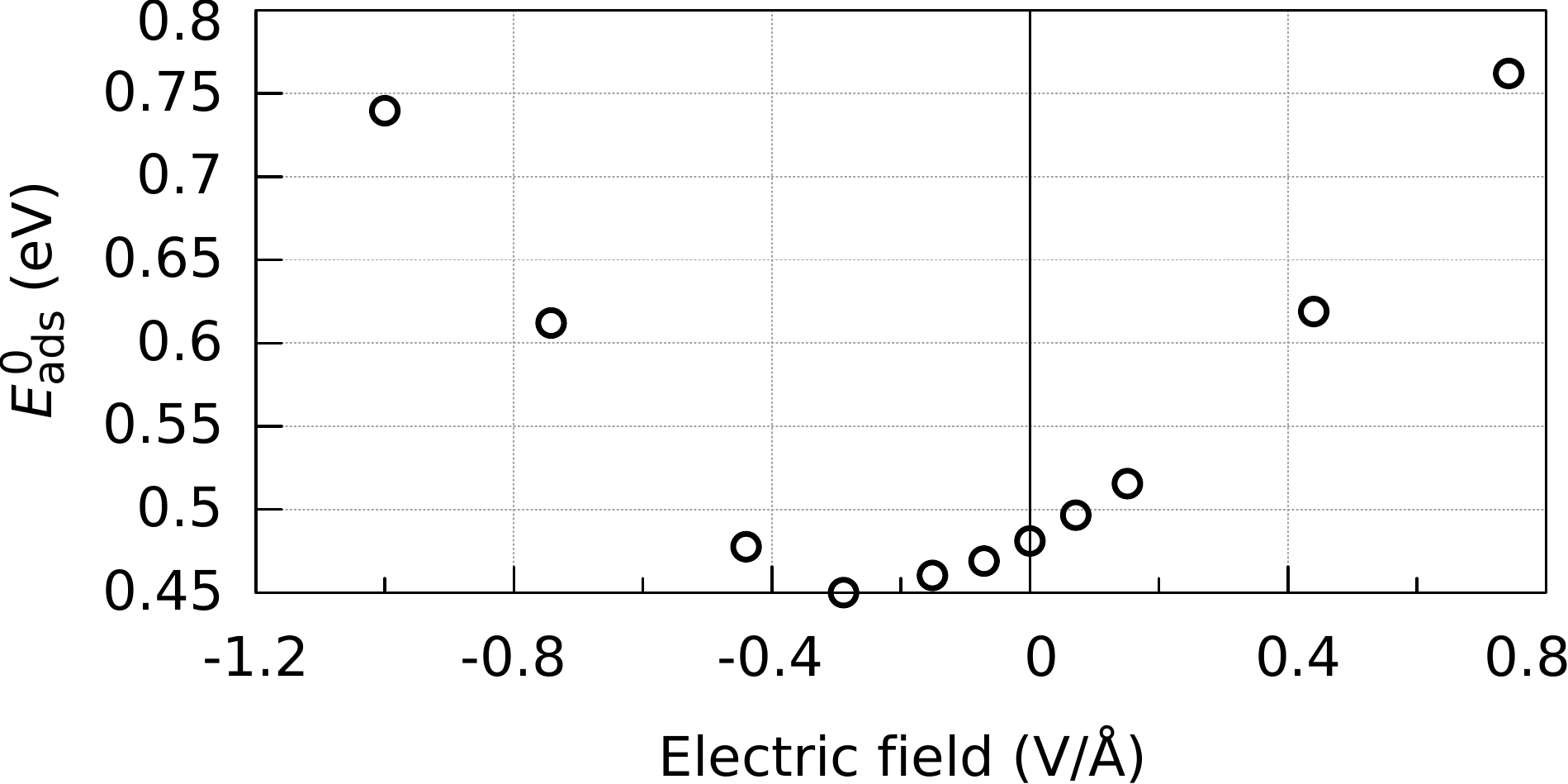}
    \caption{\rev{Adsorption energies $E^0_{\rm{ads}}$ calculated by the Expression~\ref{eq_Eads_0} at different electric field strengths.}}
    \label{fig_Eads_0}
\end{figure}

\rev{In the main text, we report the adsorption energy $E_{\rm{ads}}$ calculated as
\begin{equation}
    E_{\rm{ads}} = E_{\rm{isol}} + E_{\rm{clean}} - E_{\rm{full}},
\end{equation}
where $E_{\rm{isol}}$, $E_{\rm{clean}}$ and $E_{\rm{full}}$ denote the energies of an isolated water molecule, clean Pd surface and the surface with a water molecule adsorbed, respectively, all three relaxed in the electric fields of the same strength.
In addition to that, we report here the values of $E^0_{\rm{ads}}$ 
calculated against the zero-field gas-phase reference as follows:
\begin{equation} \label{eq_Eads_0}
    E^0_{\rm{ads}} = E_{\rm{isol}}^0 + E_{\rm{clean}} - E_{\rm{full}},
\end{equation}
where $E_{\rm{clean}}$ and $E_{\rm{full}}$ are as defined before, while $E_{\rm{isol}}^0$ is the energy of an isolated molecule with no external field applied. 
This data is given in Table~\ref{tab_Eads_0} and Fig.~\ref{fig_Eads_0}.}
\begin{table}[h]
    \centering
    \caption{\rev{ Adsorption energies $E^0_{\rm{ads}}$ at different electric field strengths relative to the zero-field reference.}} \label{tab_Eads_0}
    \setlength\tabcolsep{5 pt}
    \rev{\begin{tabular}{l|r|r|r|r|r|r|r|r|r|r}
    El. field (V/\r{A}) &
    -0.74 & -0.44 & -0.29 & -0.15 & -0.07 & No field  & 0.07  & 0.15  & 0.44  & 0.74  \\ \hline
    $E^0_{\rm{ads}}$ (eV) &
    0.612 & 0.478 & 0.450 & 0.460 & 0.469 & 0.481 & 0.497 & 0.515 & 0.618 & 0.762
    \end{tabular}
    }
\end{table}



\section{Mode-resolved ZPE effect on the dissociation barrier}

We show the contributions to the ZPE from individual vibrational modes of the adsorbed water molecule in table~\ref{tab_zpe_mode_resolved}.
\begin{sidewaystable}[h]
    \centering
    \caption{The frequencies of the individual vibrational modes of the water molecule adsorbed at Pd(111) surface (``ini") and of the transition state of its dissociation (``TS").
    The values are calculated with \gls*{pbe} + vdW$^{\rm{surf}}$.}
    \label{tab_zpe_mode_resolved}
    \setlength\tabcolsep{5 pt}
\rev{\begin{tabular}{|rr|rr|rr|rr|rr|rr|rr|rr|}
\multicolumn{2}{|c|}{-0.74} & \multicolumn{2}{c|}{-0.44} & \multicolumn{2}{c|}{-0.29} & \multicolumn{2}{c|}{-0.15} & \multicolumn{2}{c|}{0} & \multicolumn{2}{c|}{+0.15} & \multicolumn{2}{c|}{+0.44} & \multicolumn{2}{c|}{+0.74} \\
\hline
\multicolumn{1}{|c}{ini} & \multicolumn{1}{c|}{TS} & \multicolumn{1}{c}{ini} & \multicolumn{1}{c|}{TS} & \multicolumn{1}{c}{ini} & \multicolumn{1}{c|}{TS} & \multicolumn{1}{c}{ini} & \multicolumn{1}{c|}{TS} & \multicolumn{1}{c}{ini} & \multicolumn{1}{c|}{TS} & \multicolumn{1}{c}{ini} & \multicolumn{1}{c|}{TS} & \multicolumn{1}{c}{ini} & \multicolumn{1}{c|}{TS} & \multicolumn{1}{c}{ini} & \multicolumn{1}{c|}{TS} \\
\hline
30.0 & -800.5 & 27.0 & -729.5 & 29.9 & -640.5 & 24.7 & -673.1 & 13.9 & -645.4 & -23.4 & -621.7 & -41.9 & -564.4 & -42.6 & -451.7 \\
35.5 & 115.2 & 31.6 & 121.0 & 71.1 & 118.8 & 76.0 & 119.3 & 75.3 & 117.7 & 70.5 & 114.4 & 59.4 & 107.4 & 63.8 & 96.8 \\
106.2 & 167.5 & 103.2 & 180.4 & 115.9 & 183.8 & 107.7 & 180.5 & 97.9 & 183.4 & 78.5 & 186.2 & 60.5 & 183.4 & 70.8 & 170.1 \\
125.2 & 309.2 & 131.4 & 369.1 & 129.9 & 386.8 & 158.8 & 389.4 & 172.9 & 393.5 & 189.9 & 402.6 & 202.2 & 413.5 & 205.6 & 414.1 \\
262.6 & 356.7 & 196.9 & 396.4 & 416.1 & 413.6 & 437.9 & 417.0 & 449.2 & 429.3 & 457.8 & 434.5 & 409.4 & 449.3 & 378.0 & 457.4 \\
309.4 & 602.6 & 219.3 & 645.5 & 470.6 & 643.6 & 478.8 & 642.9 & 473.6 & 646.7 & 463.4 & 653.7 & 462.2 & 673.3 & 468.0 & 722.8 \\
1611.5 & 725.5 & 1593.7 & 736.9 & 1569.2 & 727.2 & 1569.1 & 732.9 & 1573.1 & 737.0 & 1576.4 & 761.1 & 1582.1 & 747.4 & 1592.0 & 751.1 \\
3597.5 & 791.1 & 3604.3 & 774.4 & 3561.5 & 773.8 & 3579.4 & 774.6 & 3600.5 & 773.9 & 3618.5 & 783.0 & 3652.5 & 768.1 & 3665.8 & 832.4 \\
3662.4 & 3605.4 & 3683.3 & 3620.3 & 3655.4 & 3624.1 & 3676.2 & 3626.5 & 3700.3 & 3628.5 & 3721.5 & 3634.5 & 3760.0 & 3632.9 & 3769.8 & 3623.6
\end{tabular}}  
\end{sidewaystable}

\begin{figure}
    \begin{center}
        \includegraphics[width=1.\textwidth]{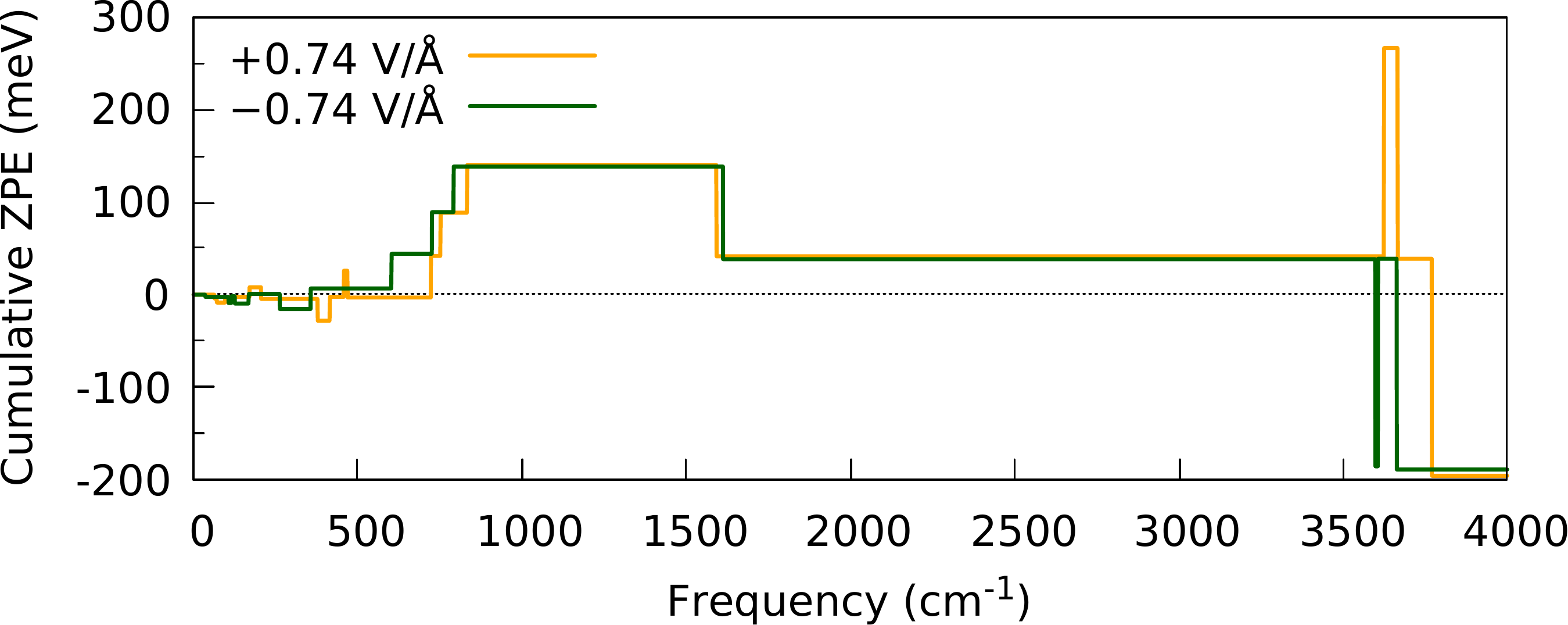}
        \caption{Cumulative contribution of the vibrational modes to the total \gls*{zpe} effect on the barrier of water dissociation, calculated by \rev{Eq.}~\ref{eq_cumulative_zpe}.}
        \label{fig_cumulative_zpe}
    \end{center}
\end{figure}

{In order to see more clearly how the shifts of the vibrational modes contribute to the total \gls*{zpe} effect on the reaction barrier, we calculate a cumulative \gls*{zpe} difference between the reactant and the transition state as follows:
\begin{equation} \label{eq_cumulative_zpe}
    \Delta^\mathrm{ZPE}_{\mathrm{cumul}}(\omega) = \sum_{\omega_i \leq \omega} \frac{\hbar \omega_i^{\rm{TS}}}{2} - \sum_{\omega_i \leq \omega} \frac{\hbar \omega_i^{\rm{ini}}}{2},
\end{equation}
where $\omega_i^{\rm{TS}}$ and $\omega_i^{\rm{ini}}$ are the normal modes of the transition state and the reactant, respectively.
We show $\Delta^\mathrm{ZPE}_{\mathrm{cumul}}(\omega)$ in figure~\ref{fig_cumulative_zpe}.}

{One can see that despite multiple shifts of the individual modes, all the \gls*{zpe} differences between the most distant electric field values of + and --0.74~V/{\AA} below 1000 cm$^{-1}$ cancel out, ending up with almost equal \gls*{zpe} contribution to the barrier.}

\section{Estimate of Tunneling Crossover Temperatures}

We estimate tunneling crossover temperature as $T_c = \hbar \omega^{\rm{TS}} / (2\pi k_B$), where $\omega^{\rm{TS}}$ is the imaginary frequency of the unstable mode at the transition state. The results are given in Table II of the main text. \rev{For comparison, in Ref.~\cite{Cao_2006_water_dissoc_CINEB}, a value of $T_c$=142 K for the dissociation of water on Pd(111) was reported, with a different exchange-correlation functional.}


\bibliography{literature}